\documentclass[prl, letterpaper, amsmath, amssymb, preprintnumbers, showpacs, superscriptaddress, twocolumn, floatfix]{revtex4-1}
\usepackage{graphicx}
\usepackage[usenames]{color}
\pdfoutput=1

\newcommand{\be}{\ensuremath{\beta} }
\newcommand{\ga}{\ensuremath{\gamma} }
\newcommand{\Ga}{\ensuremath{\Gamma} }
\newcommand{\psibar}{\ensuremath{\overline\psi} }
\newcommand{\X}{\ensuremath{\!\times\!} }
\newcommand{\lsim}{\ensuremath{\lesssim} }

\newcommand{\eq}[1]{Eq.~(\ref{#1})}
\newcommand{\fig}[1]{Fig.~\ref{#1}}
\newcommand{\refcite}[1]{Ref.~\cite{#1}}

\begin{document}
\title{Strongly interacting dynamics and the search for new physics at the LHC}

\author{T.~Appelquist}
\affiliation{Department of Physics, Sloane Laboratory, Yale University, New Haven, Connecticut 06520, USA}
\author{R.~C.~Brower}
\affiliation{Department of Physics and Center for Computational Science, Boston University, Boston, Massachusetts 02215, USA}
\affiliation{Kavli Institute for Theoretical Physics, University of California, Santa Barbara, California 93106, USA}
\author{G.~T.~Fleming}
\affiliation{Department of Physics, Sloane Laboratory, Yale University, New Haven, Connecticut 06520, USA}
\affiliation{Kavli Institute for Theoretical Physics, University of California, Santa Barbara, California 93106, USA}
\author{A.~Hasenfratz}
\affiliation{Department of Physics, University of Colorado, Boulder, Colorado 80309, USA}
\affiliation{Kavli Institute for Theoretical Physics, University of California, Santa Barbara, California 93106, USA}
\author{X.~Y.~Jin}
\affiliation{Leadership Computing Facility, Argonne National Laboratory, Argonne, Illinois 60439, USA}
\author{J.~Kiskis}
\affiliation{Department of Physics, University of California, Davis, California 95616, USA}
\author{E.~T.~Neil}
\affiliation{Department of Physics, University of Colorado, Boulder, Colorado 80309, USA}
\affiliation{RIKEN--BNL Research Center, Brookhaven National Laboratory, Upton, New York 11973, USA}
\affiliation{Kavli Institute for Theoretical Physics, University of California, Santa Barbara, California 93106, USA}
\author{J.~C.~Osborn}
\affiliation{Leadership Computing Facility, Argonne National Laboratory, Argonne, Illinois 60439, USA}
\affiliation{Kavli Institute for Theoretical Physics, University of California, Santa Barbara, California 93106, USA}
\author{C.~Rebbi}
\affiliation{Department of Physics and Center for Computational Science, Boston University, Boston, Massachusetts 02215, USA}
\author{E.~Rinaldi}
\affiliation{Nuclear and Chemical Sciences Division, Lawrence Livermore National Laboratory, Livermore, California 94550, USA}
\affiliation{Kavli Institute for Theoretical Physics, University of California, Santa Barbara, California 93106, USA}
\author{D.~Schaich}
\affiliation{Department of Physics, Syracuse University, Syracuse, New York 13244, USA}
\affiliation{Kavli Institute for Theoretical Physics, University of California, Santa Barbara, California 93106, USA}
\affiliation{Institut f\"ur Physik, Humboldt-Universit\"at zu Berlin, 12489 Berlin, Germany}
\author{P.~Vranas}
\affiliation{Nuclear and Chemical Sciences Division, Lawrence Livermore National Laboratory, Livermore, California 94550, USA}
\author{E.~Weinberg}
\affiliation{Department of Physics, Boston University, Boston, Massachusetts 02215, USA}
\author{O.~Witzel}
\affiliation{Higgs Centre for Theoretical Physics, School of Physics and Astronomy, University of Edinburgh, Edinburgh EH9 3JZ, UK}
\affiliation{Kavli Institute for Theoretical Physics, University of California, Santa Barbara, California 93106, USA}
\collaboration{Lattice Strong Dynamics (LSD) Collaboration}
\noaffiliation

\date{28 January 2016}

\preprint{EDINBURGH 2016/01; LLNL-JRNL-680732, NSF-KITP-16-004}

\begin{abstract}
We present results for the spectrum of a strongly interacting SU(3) gauge theory with $N_f = 8$ light fermions in the fundamental representation.
Carrying out non-perturbative lattice calculations at the lightest masses and largest volumes considered to date, we confirm the existence of a remarkably light singlet scalar particle.
We explore the rich resonance spectrum of the 8-flavor theory in the context of the search for new physics beyond the standard model at the Large Hadron Collider (LHC).
Connecting our results to models of dynamical electroweak symmetry breaking, we estimate the vector resonance mass to be about 2~TeV with a width of roughly 450~GeV, and predict additional resonances with masses below $\sim$3~TeV.
\end{abstract}


\maketitle

{\bf \textit{Introduction:}}
Electroweak symmetry breaking through new strong dynamics provides a potential mechanism to produce a composite scalar particle consistent with the Higgs boson discovered at the LHC~\cite{Chatrchyan:2013lba, Aad:2013wqa}.
Non-perturbative lattice calculations are a crucial tool to study relevant strongly interacting gauge theories, which must differ qualitatively from quantum chromodynamics (QCD) in order to remain phenomenologically viable.
In recent years lattice investigations have begun to explore novel near-conformal strong dynamics that emerge upon enlarging the light fermion content of such systems.
Of particular significance is increasing evidence from this work~\cite{Aoki:2013zsa, Aoki:2014oha, Athenodorou:2014eua, Fodor:2015vwa, Rinaldi:2015axa, Brower:2015owo, DelDebbio:2015byq} that such near-conformal dynamics might generically give rise to scalar ($0^{++}$) Higgs candidates far lighter than the analogous $f_0$ meson of QCD.
(See also the recent review~\cite{DeGrand:2015zxa} and references therein.)

A straightforward way to enlarge the fermion content is to increase the number $N_f$ of light fermions transforming under the fundamental representation of the gauge group SU(3).
Previous lattice studies have identified the case of $N_f = 8$ as a system that exhibits several features quite distinct from QCD, which make it a particularly interesting representative of the broader class of near-conformal gauge theories.
These features include slow running of the gauge coupling (a small \be function)~\cite{Hasenfratz:2014rna, Fodor:2015baa}, a reduced electroweak $S$ parameter~\cite{Appelquist:2014zsa}, a slowly evolving mass anomalous dimension $\ga_m$~\cite{Cheng:2013eu}, and changes to the composite spectrum including a light $0^{++}$ scalar~\cite{Aoki:2013xza, Schaich:2013eba, Aoki:2014oha, Appelquist:2014zsa}.
Although Refs.~\cite{Ishikawa:2015iwa, daSilva:2015vna, Noaki:2015xpx} even argue that the 8-flavor theory may flow to a chirally symmetric IR fixed point in the massless chiral limit, we support the conventional wisdom that chiral symmetry appears to break spontaneously for $N_f = 8$~\cite{Aoki:2013xza, Appelquist:2014zsa, Hasenfratz:2014rna, Fodor:2015baa, Schaich:2015psa, DeGrand:2015zxa}.
The 8-flavor theory continues to be investigated by several lattice groups in order to learn more about its low-energy dynamics and relate it to phenomenological model building.

Here we summarize the main results from our lattice calculations of the spectrum of the 8-flavor theory, highlighting the growing evidence for a light singlet scalar $0^{++}$ state.
We also determine the vector ($1^{--}$) and axial-vector ($1^{++}$) masses and decay constants and analyze other aspects of the rich composite spectrum of the theory, which are of phenomenological importance in the context of searches for new resonances at the LHC~\cite{ATLAS-CONF-2015-081, CMS-PAS-EXO-15-004}.
When the 8-flavor theory is responsible for electroweak symmetry breaking in models with chiral electroweak couplings assigned to only one doublet ($N_D = 1$), we estimate that the vector meson has a physical mass of about 2~TeV and a width of roughly 450~GeV.

In the context of new strong dynamics beyond the standard model, it is important for lattice calculations to be carried out using the lightest accessible fermion masses $am$, where ``$a$'' is the lattice spacing.
Small masses in turn require large lattice volumes, so we employ state-of-the-art computing~\cite{Osborn:2014kda} to investigate the lightest masses $am \geq 0.00125$ and largest volumes up to $64^3\X 128$ yet to be reached by lattice studies of the 8-flavor theory.
Our results supersede the preliminary data reported in \refcite{Rinaldi:2015axa}.
More details of our analyses will be presented in \refcite{LSD8f}.

{\bf \textit{Lattice results:}}
We investigate the spectrum in several symmetry channels for a range of light input fermion masses.
The lightest non-singlet state is a pseudoscalar meson (a $\psi\psibar$ state with quantum numbers $0^{-+}$), which we denote as $\pi$ using the corresponding QCD language for convenience.
We measure both the mass $M_{\pi}$ and the decay constant $F_{\pi}$ of this state, and we do the same for the non-singlet vector ($\rho$) and axial-vector ($a_1$) mesons.
We also investigate the singlet scalar $0^{++}$ meson, which requires computing fermion-line-disconnected contributions that we determine using U(1) stochastic sources with dilution~\cite{Rinaldi:2015axa, LSD8f}.
This disconnected calculation is currently too computationally expensive to be completed for our largest $64^3\X 128$ lattice volume.
Finally we analyze the $\psi\psi\psi$ analog of the nucleon, which we call $N$.

\begin{figure}[tbp]
  \includegraphics[width=\linewidth]{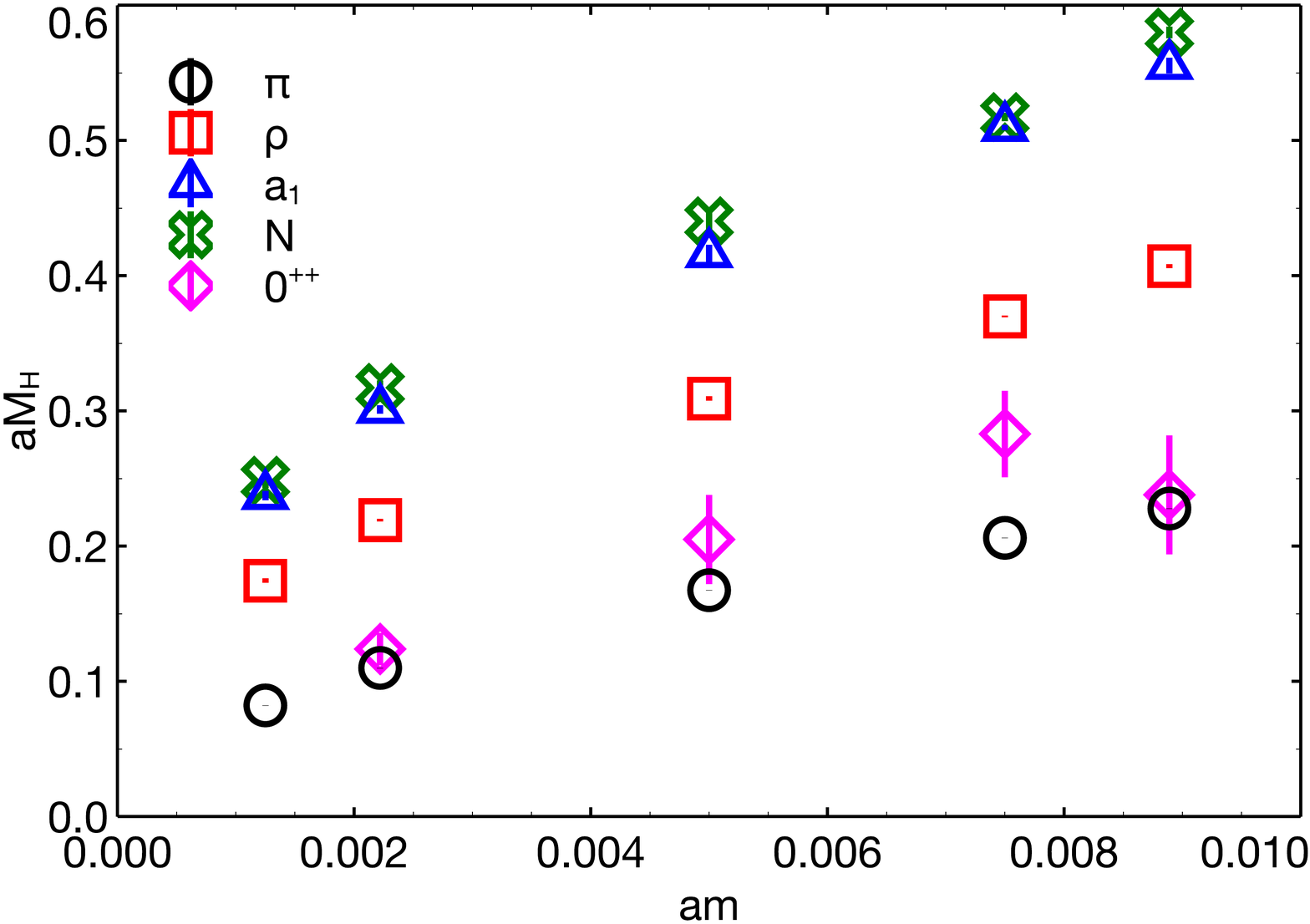}
  \caption{\label{fig:spectrum} The spectrum of the $N_f = 8$ theory, including the pseudoscalar ($\pi$), singlet scalar ($0^{++}$), vector ($\rho$), axial-vector ($a_1$) and nucleon ($N$), vs.\ the input fermion mass $am$.  The error bars, sometimes smaller than the symbols, are explained in the text.}
\end{figure}

In \fig{fig:spectrum} we plot our 8-flavor spectrum results from our largest lattice volume at each fermion mass, which ranges from $24^3\X 48$ for the heaviest $am = 0.00889$ to $64^3\X 128$ for the lightest $am = 0.00125$.
The $M_{0^{++}}$ points are weighted averages of results from fits to two different correlation functions: the dominant disconnected contribution in isolation, and the full combination of connected and disconnected contributions.
The singlet scalar error bars include systematic uncertainties from the choices of fit ranges.
These are much smaller for the other channels, where we show only statistical uncertainties.

We argue that other systematic uncertainties are also under control.
We chose $am$ to produce negligible finite-volume artifacts on the accessible lattice volumes.
Direct comparisons of the spectrum on different lattice volumes for fixed $am \geq 0.0075$ suggest that such effects are at most a few percent in our results.
We have also investigated discretization artifacts for the improved nHYP-smeared staggered lattice action~\cite{Cheng:2011ic} that we use, which we find to be similarly small~\cite{LSD8f}.

\begin{figure}[tbp]
  \includegraphics[width=\linewidth]{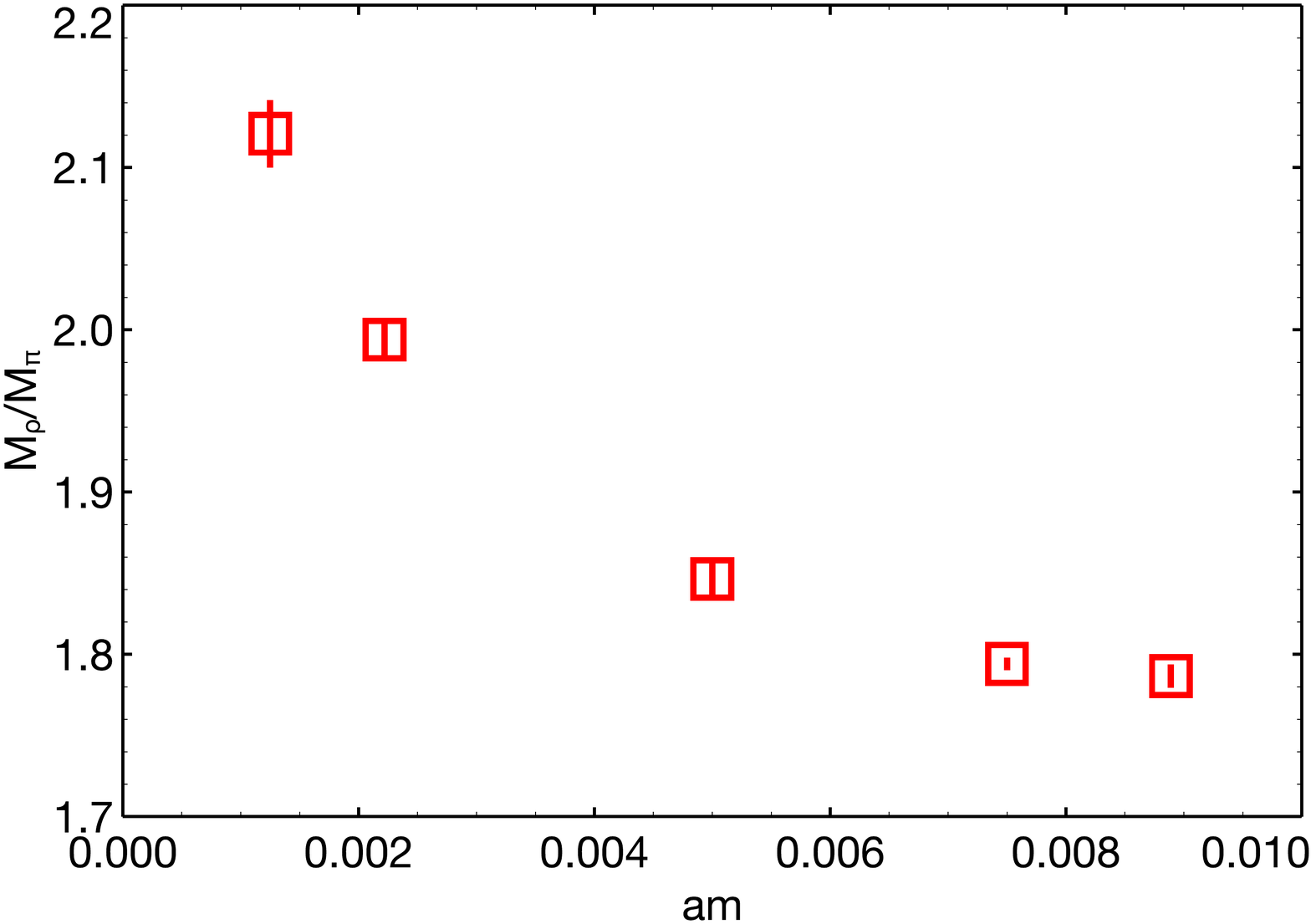}
  \caption{\label{fig:ratiovector} The ratio of the vector mass to the pseudoscalar mass vs.\ the input fermion mass $am$.}
\end{figure}

As in QCD, we find that the ratio $M_{\rho} / M_{\pi}$ steadily increases as the fermion mass decreases (\fig{fig:ratiovector}), providing further indirect evidence that the 8-flavor theory exhibits spontaneous chiral symmetry breaking.
This ratio also provides a measure of how light our fermion masses are, which can be consistently compared between different lattice studies.
Whereas previous work explored heavier mass regimes where the ratio $M_{\rho} / M_{\pi}$ was smaller ($1 \lsim M_{\rho} / M_{\pi} \lsim 1.5$ in \refcite{Aoki:2013xza}, $1.5 \lsim M_{\rho} / M_{\pi} \lsim 1.8$ in \refcite{Schaich:2013eba} and $1.25 \lsim M_{\rho} / M_{\pi} \lsim 1.45$ in \refcite{Appelquist:2014zsa}), here we reach $M_{\rho} / M_{\pi} \approx 2.1$.

Unlike QCD, the 8-flavor theory has a light singlet scalar $0^{++}$ meson with a mass $M_{0^{++}}$ that is comparable to $M_{\pi}$ in the regime we investigate.
Also, $M_{0^{++}}$ is well below both $M_{\rho}$ and $2M_{\pi}$ in this regime, which greatly simplifies the scalar spectrum analysis on the lattice: We are able to resolve the singlet scalar $0^{++}$ using the single interpolating operator $\psibar\psi$, with no need to include additional four-fermion operators or two-pseudoscalar scattering states.
Our results are notably different from QCD where the lightest singlet scalar $f_0$ is heavier than two pions.
Experimentally $m_{f_0} = 400$--550~MeV~\cite{Agashe:2014kda}, and in lattice QCD calculations with larger-than-physical pion masses this state remains as heavy as two pions and becomes heavier than $M_{\rho}$~\cite{Kunihiro:2003yj}.

\begin{figure}[tbp]
  \includegraphics[width=\linewidth]{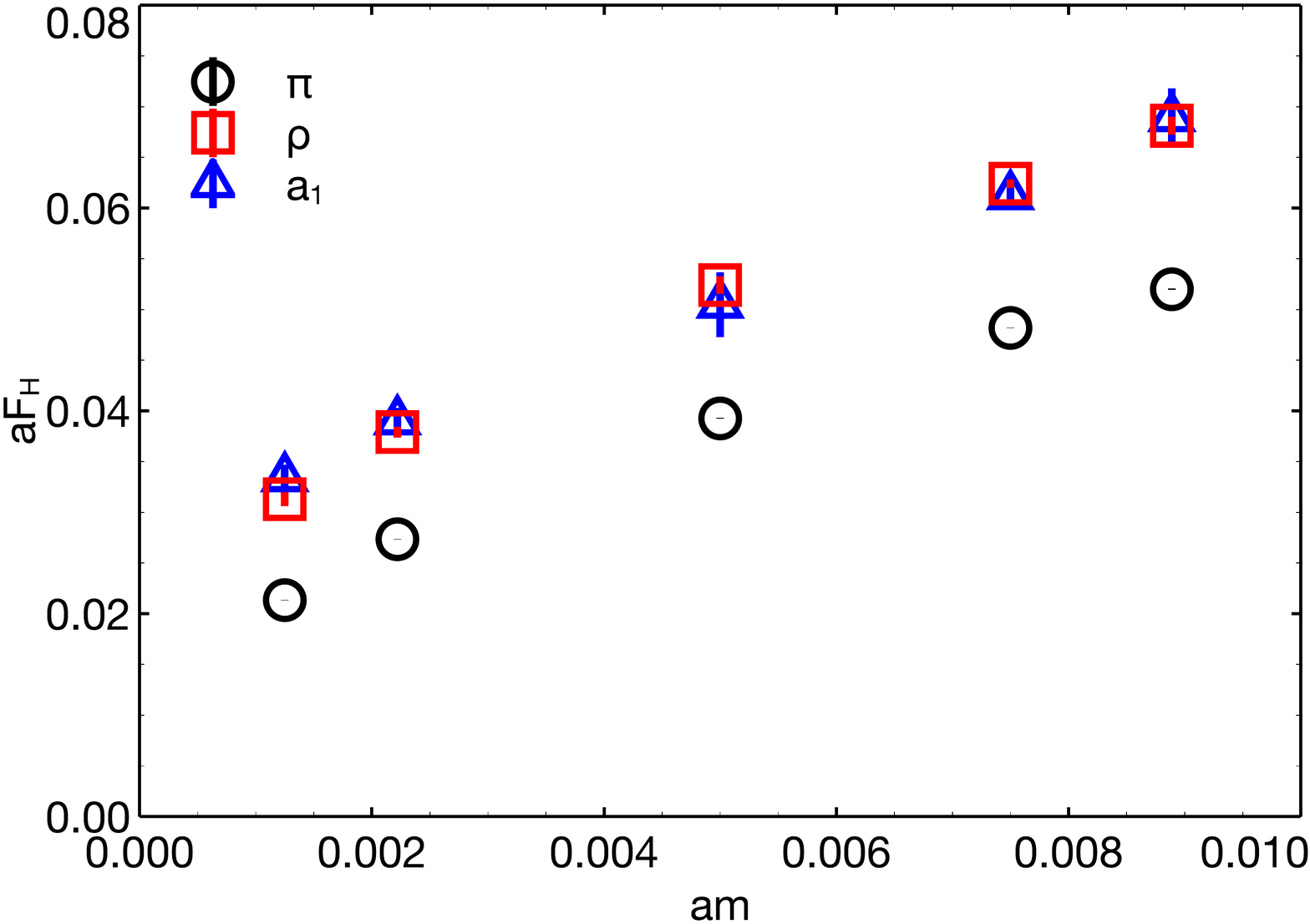}
  \caption{\label{fig:decay} The pseudoscalar, vector and axial-vector decay constants $F_{\pi}$, $F_{\rho}$ and $F_{a_1}$ vs.\ the input fermion mass $am$.  Only statistical uncertainties are shown, within which $F_{\rho} \approx F_{a_1}$ throughout the range of masses we investigate.}
\end{figure}

Our decay constant results shown in \fig{fig:decay} are also noteworthy.
The pseudoscalar decay constant $F_{\pi}$ is the smallest~\footnote{We use the ``chiral perturbation theory'' normalization of $F_{\pi}$, which corresponds to the QCD value $F_{\pi} = 92.2(1)$~MeV.  See section 5.1.1 of \refcite{Aoki:2013ldr} for a review.}, while the vector $F_{\rho}$ and axial-vector $F_{a_1}$ are approximately degenerate within statistical errors.
Similar behavior was reported in Refs.~\cite{Appelquist:2010xv, Appelquist:2014zsa}, where it was related to reductions in the electroweak $S$~parameter.

\begin{figure}[tbp]
  \includegraphics[width=\linewidth]{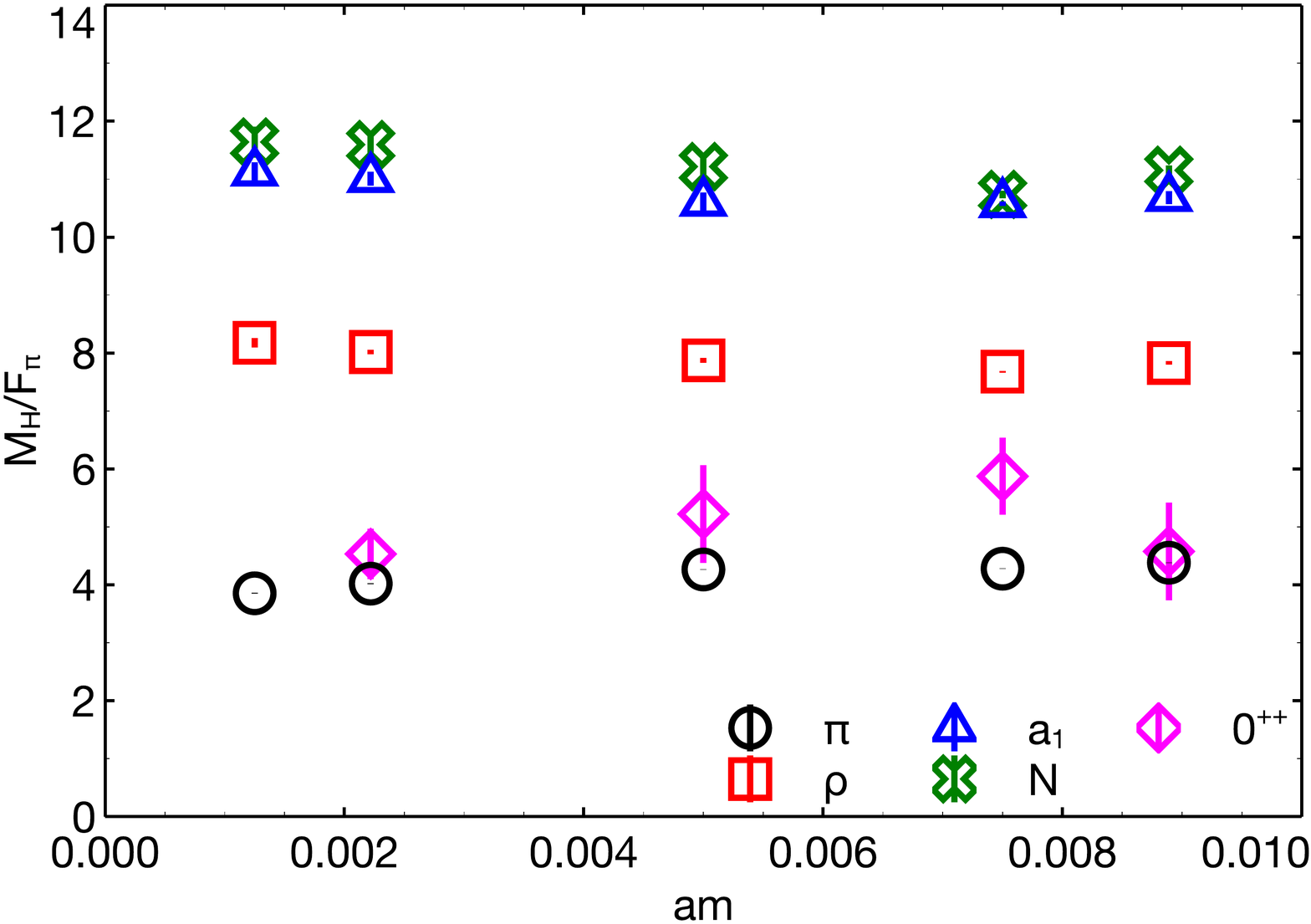}
  \caption{\label{fig:ratio} Ratios of the $N_f = 8$ hadron masses divided by the pseudoscalar decay constant $F_{\pi}$ at each fermion mass $am$.}
\end{figure}

In \fig{fig:ratio} we plot ratios of the hadron masses divided by $F_{\pi}$, observing that all the ratios are rather independent of the fermion mass, although some changes appear at our lightest mass where the vector meson is above the two-pseudoscalar threshold.
We find that $M_{\rho} / F_{\pi} \approx 8$ and $M_N / F_{\pi} \approx 11$, similar to the physical QCD ratios.
In fact, $M_{\rho} / F_{\pi} \approx 8$ appears to be a generic feature of many strongly coupled gauge theories, both IR conformal and chirally broken~\cite{Aoki:2013xza, Hayakawa:2013maa, Arthur:2014lma, Brower:2015owo, Fodor:2016wal}.

Let us now specialize to models in which we assign chiral electroweak couplings to only $N_D = 1$ pair of the $N_f = 8$ fermions.
This choice sets the physical value of $F = 246~\textrm{GeV} / \sqrt{N_D}$, and is motivated to keep the electroweak $S$~parameter as close as possible to its small experimental value~\cite{Appelquist:2010xv, Appelquist:2014zsa}.
Translating our results into physical units by identifying $F_{\pi}$ with the low-energy constant $F$ is strictly correct only in the chiral limit. 
We cannot currently carry out a controlled chiral extrapolation, in part because the effects of a light $0^{++}$ scalar on the low-energy effective theory are not yet well understood despite ongoing investigations~\cite{Soto:2011ap, Brivio:2013pma, Matsuzaki:2013eva, Buchalla:2014eca, Guo:2015isa}.
If we assume that the ratio $M_{\rho} / F_{\pi}$ shown in \fig{fig:ratio} remains relatively insensitive to the fermion masses then we would end up with a vector meson mass around 2~TeV.
Similar considerations suggest that $M_N$ and $M_{a_1}$ would be around 2.7~TeV.
On the other hand, the physical $0^{++}$ mass will depend sensitively on how long this state continues to track the pseudoscalar whose mass must vanish in the chiral limit.
At present we can estimate $0 \lsim M_{0^{++}} \lsim 1$~TeV, and this mass could be reduced further by interactions with the top quark in realistic models where the strong dynamics we study is coupled to the standard model~\cite{Foadi:2012bb}.

{\bf \textit{Comparison with previous work:}}
Some aspects of the $N_f = 8$ spectrum discussed above were observed in earlier lattice studies using different discretizations and heavier masses~\cite{Aoki:2013xza, Schaich:2013eba, Appelquist:2014zsa, Aoki:2014oha}.
In particular, the remarkably light singlet scalar $0^{++}$ Higgs candidate was first reported by \refcite{Aoki:2014oha}.
The increasing evidence~\cite{Aoki:2013zsa, Aoki:2014oha, Athenodorou:2014eua, Fodor:2015vwa, Rinaldi:2015axa, Brower:2015owo, DelDebbio:2015byq} that such behavior could be a fairly generic feature of near-conformal strong dynamics is extremely interesting from the phenomenological point of view.

\begin{figure}[tbp]
  \includegraphics[width=\linewidth]{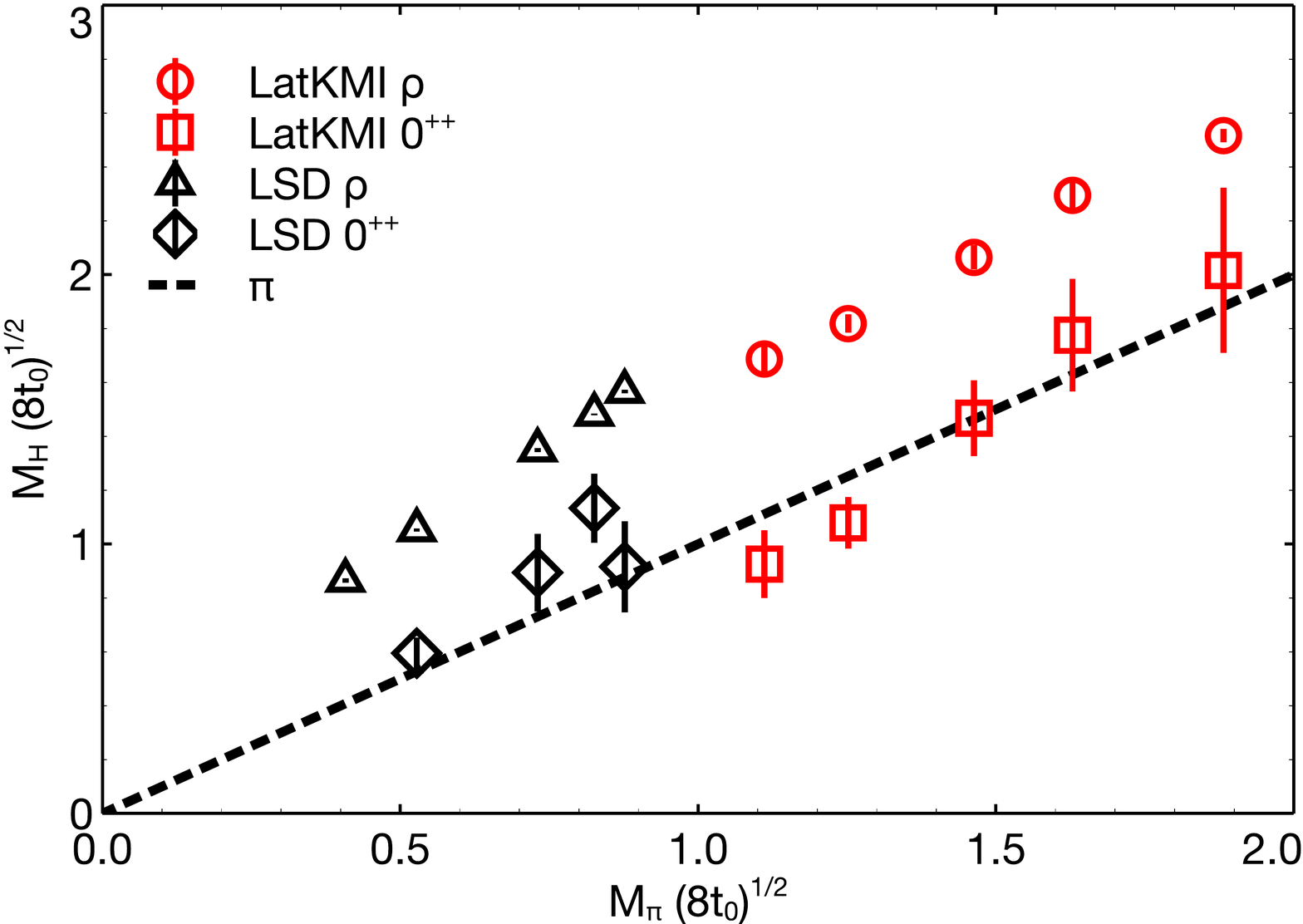}
  \caption{\label{fig:latkmi} Comparing our 8-flavor $M_{0^{++}}$ and $M_{\rho}$ results with those of the LatKMI Collaboration~\protect\cite{Aoki:2013xza, Aoki:2014oha, Aoki:2015aqa}, using the same reference scale $\sqrt{8t_0}$.  We plot these quantities vs.\ $M_{\pi}$ and include a dashed line to highlight degeneracy with the pseudoscalar meson.  A consistent trend is clearly visible, with the light singlet scalar $0^{++}$ state following the pseudoscalar to the smallest masses studied so far.}
\end{figure}

\begin{figure*}[tbp]
  \includegraphics[width=0.45\linewidth]{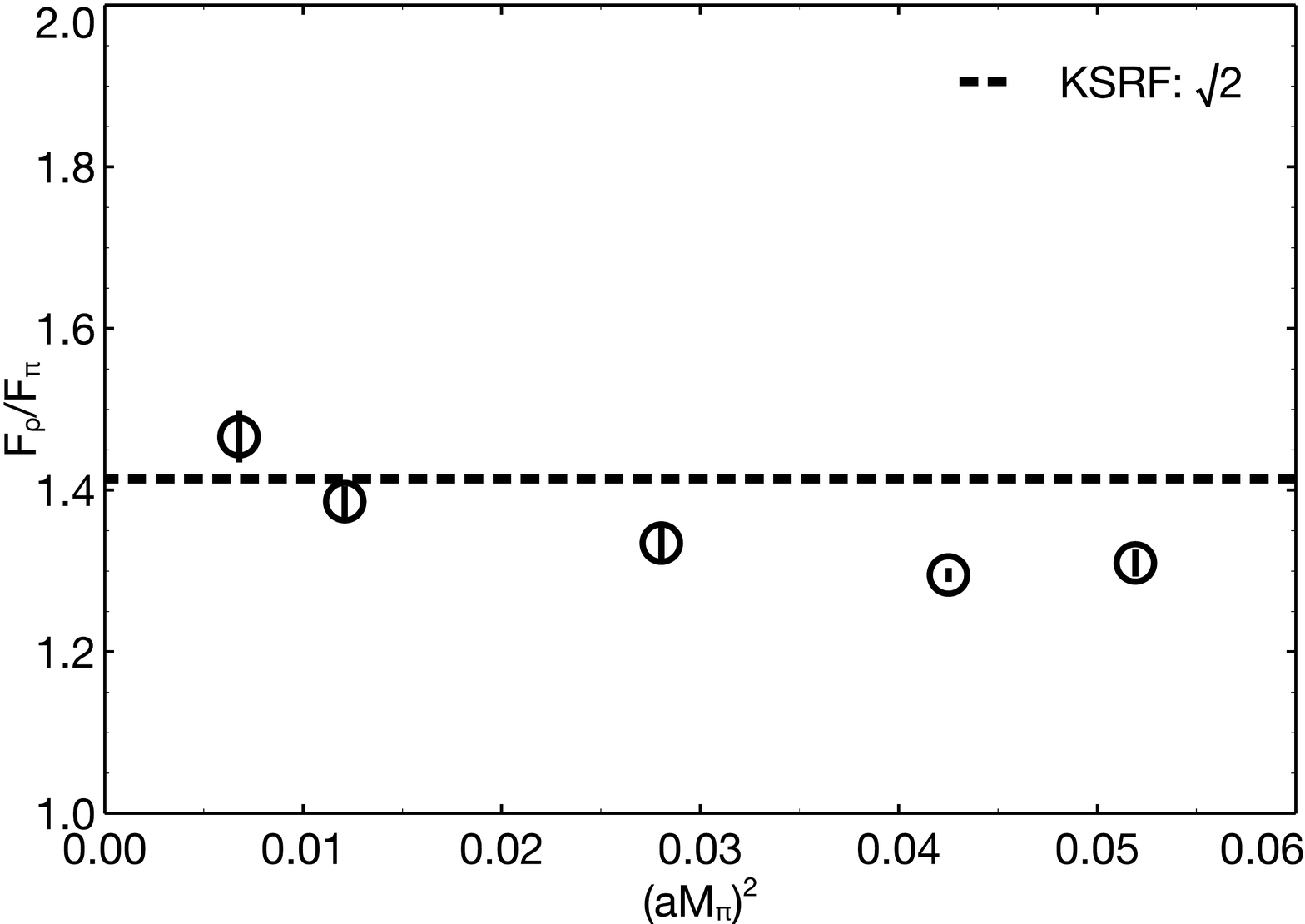}\hfill \includegraphics[width=0.45\linewidth]{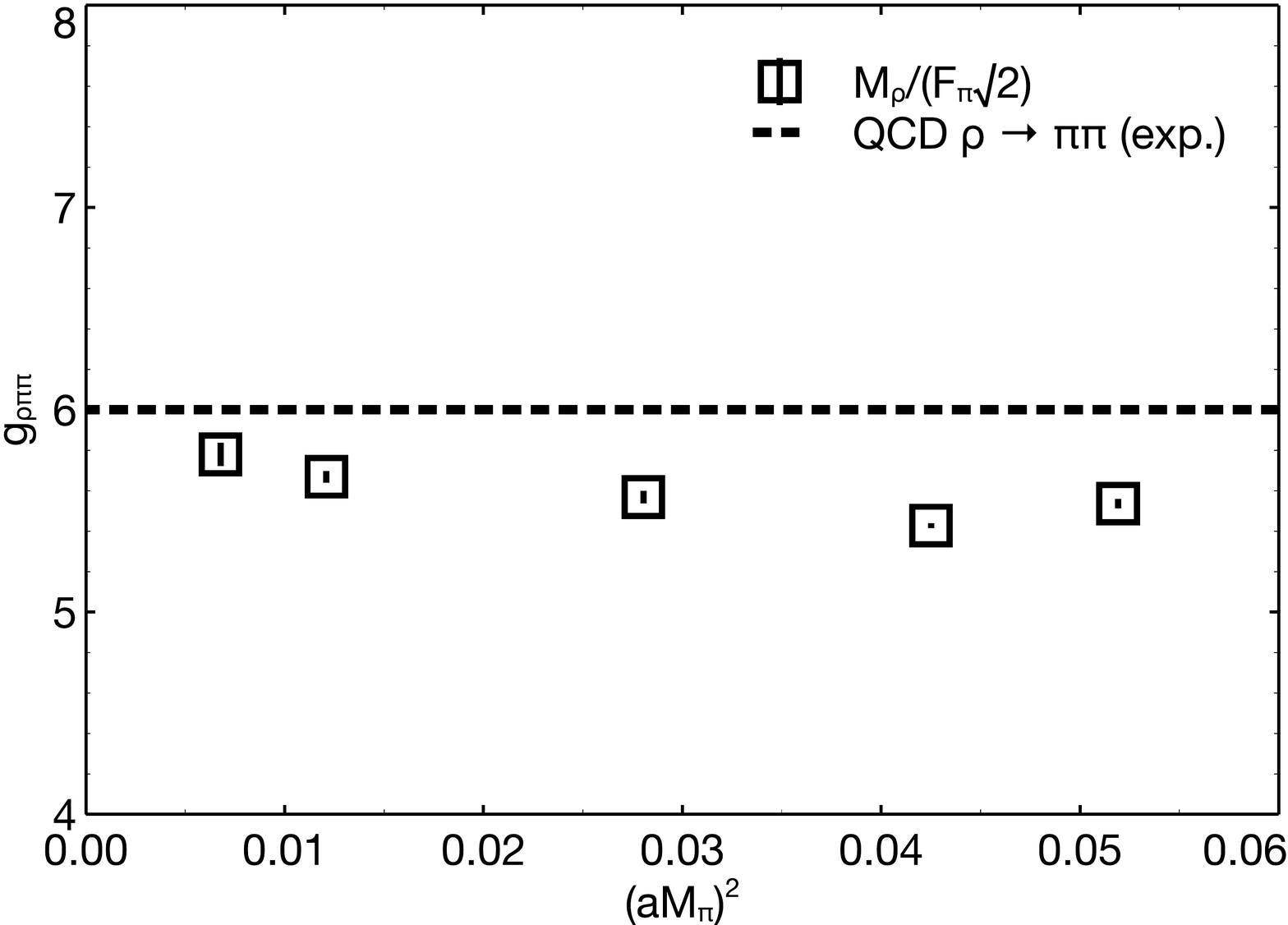}
  \caption{\label{fig:KSRF} {\bf Left:} Testing the first KSRF relation in \protect\eq{eq:KSRF} for $N_f = 8$ through lattice measurements of $F_{\rho} / F_{\pi}$. {\bf Right:} The second KSRF relation then provides an estimate for $g_{\rho\pi\pi} \approx M_{\rho} / \left(\sqrt{2} F_{\pi}\right)$, which is within 10\% of the physical QCD value $g_{\rho\pi\pi} \approx 6$ throughout the range of masses we investigate.}
\end{figure*}

In this context the behavior of $M_{0^{++}}$ in the chiral limit is particularly important.
As a step in this direction, in \fig{fig:latkmi} we compare our results for the light meson spectrum with those of the LatKMI Collaboration~\cite{Aoki:2013xza, Aoki:2014oha, Aoki:2015aqa}.
To enable consistent comparisons between independent studies that employ different lattice actions, we plot all quantities in terms of a standard Wilson flow reference scale $\sqrt{8t_0}$ introduced in \refcite{Luscher:2010iy}.
This figure demonstrates that our work accesses significantly lighter masses.
At the same time, the clear consistency between the two sets of results confirms that discretization artifacts are under control.
As discussed above, both studies find $M_{0^{++}} \approx M_{\pi} \ll M_{\rho}$, and it remains an open question how light the scalar will become in the chiral limit where $M_{\pi} \to 0$.
In the absence of non-perturbative lattice calculations it would have been difficult to anticipate this dramatic dynamical effect.

{\bf \textit{The vector meson:}}
We now study more properties of the vector resonance to further relate our numerical work to phenomenological models of new strong dynamics.
The production rate of the vector meson at colliders is determined by its couplings to standard model fermions, which are in turn related to the decay constant $F_{\rho}$.
On the other hand, the resonance's decay rate is dominated by its coupling to the longitudinal components of the electroweak gauge bosons, assuming that the $N_f^2 - 4 = 60$ uneaten pseudo-Nambu--Goldstone bosons are heavy enough that their effect is negligible.
The resulting decay width $\Ga_{\rho}$ of the vector resonance therefore depends on the $\rho \to \pi\pi$ coupling $g_{\rho\pi\pi}$ of the new strong dynamics.

We estimate $g_{\rho\pi\pi}$ invoking the Kawarabayashi--Suzuki--Riazuddin--Fayyazuddin (KSRF) relations~\cite{Kawarabayashi:1966kd, Riazuddin:1966sw}
\begin{align}
  \label{eq:KSRF}
  F_{\rho} & = \sqrt{2} F_{\pi} &
  g_{\rho\pi\pi} & =  \frac{M_{\rho}}{\sqrt{2} F_{\pi}},
\end{align}
in a manner similar to what has been done in lattice QCD studies such as \refcite{Jansen:2009hr}.
These relations result from fairly simple assumptions (principally current algebra and some form of vector meson dominance~\cite{Johnson:1970xe}), and arise rather generically in models of hidden local symmetries~\cite{Bando:1987br} and in chiral effective theories for spin-1 mesons~\cite{Birse:1996hd}.
We assess their applicability to $N_f = 8$ through our direct measurements of $F_{\rho}$ and $F_{\pi}$.

In the left panel of \fig{fig:KSRF} we plot our lattice results for $F_{\rho} / F_{\pi}$, finding agreement with the first KSRF relation in \eq{eq:KSRF} to within 8\% throughout the range of masses we investigate.
This justifies using the second KSRF relation to estimate $g_{\rho\pi\pi} \approx M_{\rho} / \left(\sqrt{2} F_{\pi}\right)$.
When we plot this quantity in the right panel of \fig{fig:KSRF}, we observe that it is within 10\% of the QCD value $g_{\rho\pi\pi} \approx 6$.
Since we have already seen that the 8-flavor $M_{\rho} / F_{\pi} \approx 8$ is similar to the QCD value and rather independent of the fermion mass, this behavior is not too surprising.

The physical decay width of the vector resonance can now be estimated as 
\begin{equation}
  \Ga_{\rho} \approx \frac{g_{\rho\pi\pi}^2 M_{\rho}}{48\pi} \approx \frac{M_{\rho}^3}{96\pi F_{\pi}^2}.
\end{equation}
Here we neglect the small electroweak gauge boson masses compared to the vector resonance mass.
With $M_{\rho} \simeq 2$~TeV, this expression leads to $\Ga_{\rho} \simeq 450$~GeV. 
The corresponding $\Ga_{\rho} / M_{\rho} \simeq 0.22$ for $N_f = 8$ is also similar to the QCD value, 0.19~\cite{Agashe:2014kda}.
This relatively broad width may make such a vector resonance challenging to discover at the LHC~\cite{Murayama:2014yja}.

It is significant that we are able to measure $F_{\rho}$ and estimate $g_{\rho\pi\pi}$ and $\Ga_{\rho}$ using lattice calculations and the KSRF relations.
These quantities are needed for phenomenological predictions of vector meson production and decay rates at colliders in models of new strong dynamics such as those considered by Refs.~\cite{Fukano:2015hga, Franzosi:2015zra}.
In particular, \refcite{Fukano:2015hga} needs to treat $F_{\rho}$ and $g_{\rho\pi\pi}$ as tunable parameters.
Our new non-perturbative results for these quantities may be used to improve this aspect of strongly interacting model building in the future.

{\bf \textit{Conclusions:}}
Theories with near-conformal strong dynamics can provide a phenomenologically viable Higgs candidate and can be investigated on the lattice.
Through lattice studies of a representative system, strongly interacting 8-flavor SU(3) gauge theory, we have shown the dynamical generation of a light singlet scalar $0^{++}$ state that is roughly degenerate with the pseudoscalar down to the lightest masses that have yet been considered.
This behavior contrasts with QCD and leaves open the possibility that this $0^{++}$ could provide the 125~GeV Higgs boson in full phenomenological models based on near-conformal strong dynamics.

In addition we investigated heavier states of the 8-flavor theory and estimated physical resonance masses in the energy range currently being explored at the LHC.
These predictions resulted from ratios such as $M_{\rho} / F_{\pi}$, which can be curiously similar to their counterparts in QCD and show only mild dependence on the fermion mass throughout the regime we can access using state-of-the-art computing.
In addition to the $M_{\rho} \simeq 2$~TeV vector meson, a plethora of scalar, pseudoscalar and $2^{++}$ glueball resonances also exist, some of which may be light enough to explain tentative excesses observed experimentally~\cite{ATLAS-CONF-2015-081, CMS-PAS-EXO-15-004}.

We also studied other properties of the vector resonance, including its decay constant $F_{\rho}$ and the $g_{\rho\pi\pi}$ coupling, which are often needed by phenomenological models of new strong dynamics.
In addition to observing $F_{\rho} \approx F_{a_1}$, we found that our lattice results for $F_{\rho}$ and $F_{\pi}$ satisfy the first KSRF relation.
This motivated us to use the second KSRF relation to estimate $g_{\rho\pi\pi}$, which we also found to be similar to its QCD counterpart.
Our lattice results and the KSRF relations finally produce an estimated vector width of $\Ga_{\rho} \simeq 450$~GeV, a relatively large value that may be challenging to resolve at the LHC.

We expect a renewed interest in new strong dynamics beyond the standard model, and we have highlighted some of the results that can now be obtained from numerical lattice calculations.
The results reported here are a significant step towards a more complete understanding of near-conformal strongly coupled theories, which can help guide model building in the search for new resonances.
To establish further connections between lattice studies and phenomenology we plan to investigate the scattering of the light $0^{++}$ scalar and the pseudoscalar, which will provide non-perturbative information about the relevant low-energy effective theory~\cite{Soto:2011ap, Brivio:2013pma, Matsuzaki:2013eva, Buchalla:2014eca, Guo:2015isa}.

\vspace{12 pt}
\noindent {\sc Acknowledgments:}~We thank the LatKMI Collaboration for sharing their $\sqrt{8t_0}$ results prior to the publication of \refcite{Aoki:2015aqa}.
We thank the Lawrence Livermore National Laboratory (LLNL) Multiprogrammatic and Institutional Computing program for Grand Challenge allocations and time on the LLNL BlueGene/Q supercomputer, along with funding from LDRD~13-ERD-023.
Additional numerical analyses were carried out on clusters at LLNL, Boston University and Fermilab.
Part of this work was performed at the Aspen Center for Physics (R.C.B., G.T.F., A.H., C.R.\ and D.S.) supported by the U.S.~National Science Foundation (NSF) under Grant No.~1066293, and at the Kavli Institute for Theoretical Physics (R.C.B., G.T.F., A.H., E.T.N., E.R., D.S.\ and O.W.) supported by NSF Grant No.~PHY11-25915.
D.S.\ was supported by the U.S.~Department of Energy (DOE) under Grant Nos.~{DE-SC0008669} and {DE-SC0009998}.
A.H.\ and E.T.N.\ were supported by DOE grant~{DE-SC0010005}; Brookhaven National Laboratory is supported by the DOE under contract~{DE-SC0012704}.
R.C.B., C.R.\ and E.W.\ were supported by DOE grant~{DE-SC0010025}.
In addition, R.C.B.\ and C.R.\ acknowledge the support of NSF grant OCI-0749300.
O.W.\ was supported by STFC, grant ST/L000458/1; this project has received funding from the European Union's Horizon 2020 research and innovation programme under the Marie Sk{\l}odowska-Curie grant agreement No.~659322.
G.T.F.\ was supported by NSF grant PHY-1417402.
E.R.\ and P.V.\ acknowledge the support of the DOE under contract DE-AC52-07NA27344 (LLNL).
Argonne National Laboratory is supported by the DOE under contract DE-AC02-06CH11357.

\raggedright
\bibliographystyle{apsrev}
\bibliography{KS_nHYP_8f-short}
\end{document}